\documentclass[10pt,a4paper,twocolumn,hidelinks]{article}

\usepackage{comment}

\usepackage{latexsym}      % needed math symbols
\usepackage{hyperref}
\hypersetup{
 colorlinks=true,
 citecolor=blue,
 linkcolor=blue,
 urlcolor=blue,
 pdfpagemode=UseNone,
 pdfstartview=FitH}
\usepackage{graphicx}      % for importing eps figures
\usepackage{amsmath}       % for advanced math symbols
\usepackage[affil-it]{authblk} 			% found in preprint bundle package, http://ctan.org/pkg/preprint
\usepackage[margin=2cm]{geometry} % paper and margin formats as set by SPP

\usepackage{sectsty}

\makeatletter
\newcommand{\Capitalize}[1]{%
  \edef\@tempa{\expandafter\@gobble\string#1}%
  \edef\@tempb{\expandafter\@car\@tempa\@nil}%
  \edef\@tempa{\expandafter\@cdr\@tempa\@nil}%
  \uppercase\expandafter{\expandafter\def\expandafter\@tempb\expandafter{\@tempb}}%
  \@namedef{\@tempb\@tempa}{\expandafter\MakeUppercase\expandafter{#1}}}
\makeatother

\makeatletter
\newbox\abstract@box
\renewenvironment{abstract}
  {\global\setbox\abstract@box=\vbox\bgroup
     \hsize=\textwidth\linewidth=\textwidth
    \small
    \begin{center}%
    {\bfseries \abstractname\vspace{-.5em}\vspace{\z@}}%
    \end{center}%
    \quotation}
  {\endquotation\egroup}
\expandafter\def\expandafter\@maketitle\expandafter{\@maketitle
  \ifvoid\abstract@box\else\unvbox\abstract@box\if@twocolumn\vskip1.5em\fi\fi}
\makeatother
\providecommand{\keywords}[1]{\noindent \mdseries{{Keywords:}} #1}
\providecommand{\DOI}[1]{\vspace{1\baselineskip}}

\makeatletter
\renewcommand\section{\@startsection
   {section}{1}{0pt}%
   {-\baselineskip}%
   {0.3\baselineskip}%
   {\normalfont\bfseries}}%
\makeatother
\renewcommand\thesection{\Roman{section}}

\makeatletter
\renewcommand\subsection{\@startsection
   {subsection}{1}{0pt}%
   {-\baselineskip}%
   {0.3\baselineskip}%
   {\normalfont\bfseries}}%
\makeatother
\renewcommand{\thesubsection}{\thesection.\arabic{subsection}}

\usepackage{booktabs}
\usepackage{dcolumn}
\newcolumntype{d}[1]{D{.}{.}{#1}}

\usepackage[labelfont=bf]{caption}
\usepackage{fancyhdr}
\fancypagestyle{titlestyle}
{

\fancyhf[c]{ }
\fancyhf[r]{ }
\fancyfoot[c]{1}
}

\usepackage{bm}

\usepackage{textcomp}
\usepackage{graphicx}
\usepackage{dcolumn}
\usepackage{bm}
\usepackage{tikz}
\usetikzlibrary{shapes.geometric, arrows}
\usetikzlibrary{positioning}
\usetikzlibrary{shapes,arrows}
\usetikzlibrary{intersections}
\usepackage{csvsimple}
\usepackage{changepage}
\usepackage[tbtags]{mathtools}
\usepackage{siunitx}
\usepackage{csquotes}
\usepackage{enumerate}
\usepackage{enumitem}
\usepackage{float}
\usepackage{amsmath}
\usepackage{wasysym}

 \usepackage[
 	nonumberlist, 				
 	acronym,
 	nomain,
 	nopostdot,
 ]{glossaries}
 \usepackage{glossary-superragged}
\usepackage{xcolor}

\usepackage{pgfplots}
\usepackage{tikz}
\usetikzlibrary{arrows}
\usepackage{amsmath}
\pgfplotsset{compat=newest}
\usepgfplotslibrary{fillbetween}
\usetikzlibrary{calc}
\def\centerarc[#1](#2)(#3:#4:#5)% Syntax: [draw options] (center) (initial angle:final angle:radius)
{ \draw[#1] ($(#2)+({#5*cos(#3)},{#5*sin(#3)})$) arc (#3:#4:#5); }
\usetikzlibrary{external}
\usepackage{blindtext}
\usepackage{amssymb}

\usepackage{nicefrac}

\usepackage{multirow}
\usepackage{tabularx}
\usepackage{booktabs}
\usepackage{array}
\usepackage{longtable}
\newcolumntype{L}[1]{>{\raggedright\let\newline\\\arraybackslash\hspace{0pt}}m{#1}}
\newcolumntype{C}[1]{>{\centering\let\newline\\\arraybackslash\hspace{0pt}}m{#1}}
\newcolumntype{R}[1]{>{\raggedleft\let\newline\\\arraybackslash\hspace{0pt}}m{#1}}

\usepackage[
    backend=biber,
    style=nature,
  ]{biblatex}

\usepackage[normalem]{ulem}
\usepackage{microtype}

\definecolor{correccolortwo}{rgb}{0.85, 0.0, 0.1}

\definecolor{deleted}{rgb}{0.66, 0.66, 0.66}

\usepackage{titlesec}
\titleformat{\section}
  {\bf\sffamily}
  {\thesection. }
  {5pt}
  {\MakeUppercase}
\renewcommand{\thesection}{\Roman{section}} 

\titleformat{\subsection}
  {\bf\sffamily}
  {\thesubsection. }
  {5pt}{}

\newcommand{\BV}[1]{\textcolor{red}{BV: #1}}

\usepackage{stackengine}
\usepackage{caption}
\usepackage{subcaption}
\usepackage{graphicx}
\usepackage{relsize}
\usepackage{cleveref}

\usepackage{stfloats} % Add this package for improved figure placement options

\DeclareRobustCommand{\orderof}{\ensuremath{\mathcal{O}}}

\begin{document}

\title{\textbf{Long-term microgravity experiments reveal a new mechanism for particle aggregation in suspension}}

\author[1]{Fabian Kleischmann\thanks{Corresponding author: fabian.kleischmann@tu-dresden.de}}
\author[1,2]{Bernhard Vowinckel}
\author[2]{Eckart Meiburg}
\author[2]{Paolo Luzzatto-Fegiz}

\affil[1]{Institute of Urban and Industrial Water Management, Technische Universit\"{a}t Dresden, 01062 Dresden, Germany}
\affil[2]{Department of Mechanical Engineering, UC Santa Barbara, Santa Barbara, CA 93106, USA}

\date{\today}

\begin{abstract}
\noindent
% 137 of 150 words
Microgravity experiments on board the International Space Station, combined with particle-resolved direct numerical simulations, were conducted to investigate the long-term flocculation behavior of clay suspensions in saline water in the absence of gravity.
After an initial homogenization of the suspensions, different clay compositions were continuously monitored for $99$ days, allowing a detailed analysis of aggregate growth through image processing.
The results indicate that the onboard oscillations \mbox{(g-jitter)} may have accelerated the aggregation process. 
Aggregate growth driven by these oscillations is found to occur at a faster rate than aggregation caused by Brownian motion.
This effect is further confirmed by numerical simulations, which also demonstrated that parameters such as the oscillation amplitude and the solid volume fraction influence growth acceleration.
These findings highlight that oscillations may act as a previously unrecognized mechanism that contributes to particle aggregation in fluids.
\linebreak


\end{abstract}

\begingroup
\sffamily
\maketitle
\endgroup

\thispagestyle{titlestyle}

%%%%%%%%%%%%%%%%%%%%%%%%%%%%%%%%%%%%%%%%
%%%%%%%%%%%%%%%%%%%%%%%%%%%%%%%%%%%%%%%%
\section*{Introduction}\label{sec:intro}
%%%%%%%%%%%%%%%%%%%%%%%%%%%%%%%%%%%%%%%%
%%%%%%%%%%%%%%%%%%%%%%%%%%%%%%%%%%%%%%%%

Particle aggregation in suspensions is a fundamental process for the behavior and transport of particulate matter in both natural and industrial settings.
It is a key element for ecological balance in nature, as it contributes to sediment transport, water purification, and the nutrient cycle of oceans, lakes, and rivers \cite{2019_Camassa_etal}. 
On the other hand, in industry, aggregation is targeted for applications in water treatment, mining, pharmaceutical production, food processing, and battery manufacturing \cite{2012_Chong, 2013_Tadros, 2011_Zhu_etal}.

The aggregation process of suspended particles consists of two parts:
first, two or more particles come very close together (accumulation) and possibly in contact (collision), and second, the particles stay together, forming a floc \cite{2004_Kim_Stolzenbach} due to surface forces in direct contact, material properties that form bridges, or interlocking due to the shape of the particles \cite{2007_Tomas}.
The physical process of bringing particles together and facilitating aggregation is attributed to three mechanisms: (i)~Brownian motion, (ii)~differential settling and (iii)~fluid shear \cite{2009_Partheniades}. 
Brownian motion is a colloidal behavior and important for small particles ($\leq 2\mu$m) in quiescent fluid, where individual entities collide with fluid molecules and undergo a direction-independent random walk \cite{2001_Baldwin_Dempsey}.
The intensity of particle movement decreases with increasing particle size and fluid viscosity.
As the length scale of the particle size increases, differential settling and liquid shear become more relevant \cite{1988_vanLeussen, 2001_Baldwin_Dempsey, 2009_Burd_Jackson}.
Differential settling describes the process of particles and aggregates settling at different velocities caused by varying particle properties such as size, density, solid volume fraction and shape \cite{1988_Lick_Lick, 2004_Kim_Stolzenbach, 2019_Vowinckel_etal, 2025_Metelkin_Vowinckel}.
This mechanism is predominant when the fluid is at rest or as long as settling is dominant for particle motion.
Once fluid flow becomes the dominant driving force of particle motion, aggregation is governed mainly by fluid shear, where particles move at different speeds due to velocity gradients in the flow field \cite{2004_Kim_Stolzenbach}.
All three mechanisms, Brownian motion, differential settling, and fluid shear, increase the probability of particle collisions with subsequent aggregation by inducing relative motion between two or more particles.

A key property of particles for aggregation is cohesion, an attractive force between particles due to electrochemical forces, which becomes an important factor for clay and silt materials with particle sizes smaller than $63 \, \mu m$ \cite{2011_Grabowski_etal}.
Cohesion is known to strengthen the bonds between particles, increasing the stability of aggregates and allowing the formation of larger structures \cite{2019_Vowinckel_etal, 2020_Zhao_etal}.
As aggregates grow, their effective mass increases, causing gravitational forces to be the dominant factor in their motion under Earth-bound conditions, rather than cohesive forces \cite{2022_Turetta_Lattuada}.
This phenomenon hinders the isolation and investigation of the long-term effects of cohesion on the aggregation behavior of particles.
Since gravity is an omnipresent phenomenon on Earth, methods such as drop towers or parabolic flights have increasingly gained attention in recent decades to investigate the characteristics of particulate suspensions in weightlessness \cite{2005_Simic-Stefani_etal}. 
However, these experiments provide weightless conditions for only a few seconds, significantly restricting the scope of investigation.

It is against this background that long-term experiments on board the International Space Station (ISS) have allowed new possibilities for investigating particle behavior in microgravity.
The facilities on board the ISS have already been used for a large number of studies to investigate crystal formation \cite{2022_Wright_etal, 2024_Jackson_etal} or colloidal aggregation \cite{2012_Veen_etal, 2023_Miki_etal}.
The conditions on board the ISS are almost weightless, as the gravitational field is several orders of magnitude weaker than on Earth.
However, it needs to be taken into account that inertia effects still exist due to the presence of onboard accelerations.
According to \cite{1996_Monti_Savino}, these accelerations can be caused by three factors:
(i) external sources, e.g. forces acting on the space station due to aerodynamics and operational procedures, 
(ii) mass ejections, due to rocket-powered activities such as docking or reboost maneuvers, and
(iii) internal causes arising from the presence of crew members and machinery, including fans, compressors, and pumps \cite{2016_McPherson_etal}.
The effects of these accelerations can be divided into constant accelerations and oscillating accelerations \cite{1996_Monti_Savino, 2001_Ramachandran, 2002_Savino}.
The latter is referred to as g-jitter and has gained a lot of attention in the recent past. 
Crystallization experiments in the first International Microgravity Laboratory (IML-1) \cite{1996_Trolinger_etal} and on board the ISS \cite{2000_Lorber_etal, 2001_Carotenuto_etal, 2001_Otalora_etal} have shown that the effects of g-jitter can lead to random particle motion in microgravity, known as an inertial random walk \cite{1991_Trolinger_etal}.
Tracing the trajectories of individual randomly moving particles showed that the effect of g-jitter causes convective flows that lead to higher particle velocities than those obtained by diffusion \cite{2006_Simic-Stefani_etal}.
These flows lead to clustering of particles due to the inertia of the particles, characterized by high particle concentrations in local regions of the surrounding fluid, generally termed inertial clustering \cite{2013_Lappa}.

The crystallization process in microgravity has been extensively studied since the first experiments in the early seventies. 
During the past decades, more than $500$ studies have explored the effects of microgravity on crystal formation.
These investigations have significantly advanced both experimental techniques and the understanding of crystal manufacturing and aggregation dynamics \cite{2022_Wright_etal, 2024_Jackson_etal}. 
This has generated valuable expertise on controlled growth processes and the influence of g-jitter, which contributes to a wide range of scientific and technical applications beyond crystallization \cite{2021_Snell_Helliwell}.
However, most studies focused mainly on manufactured materials and experimental durations of only a few days.
Although these conditions were completely sufficient for the intended applications, this opens up the questions of the extent to which the aggregate behavior of natural materials changes and how this behaves over a long period of time.

To address this, we performed microgravity experiments on board the ISS to investigate the aggregation process of natural clay particles in saline suspensions over a time period that exceeds previous campaigns by far.
The setup design allowed the exclusion of differential settling and fluid shear as driving factors for particle contact.
The impact of g-jitter was monitored and investigated in detail.
We supplement these experiments with a campaign of particle-resolved direct numerical simulations (pr-DNS) to further investigate the governing mechanisms responsible for the long-term dynamics of particle aggregation. 
This allows elucidating potential aggregation mechanisms beyond differential settling and fluid shear and to show how these mechanisms scale with varying properties of the fluid and suspended particles, respectively.

%%%%%%%%%%%%%%%%%%%%%%%%%%%%%%%%%%%%%%%%
%%%%%%%%%%%%%%%%%%%%%%%%%%%%%%%%%%%%%%%%
\section*{Methods}\label{sec:method}
%%%%%%%%%%%%%%%%%%%%%%%%%%%%%%%%%%%%%%%%
%%%%%%%%%%%%%%%%%%%%%%%%%%%%%%%%%%%%%%%%

%%%%%%%%%%%%%%%%%%%%%%%%%%%%%%%%%%%%%%%%
\subsection*{Experimental setup}
%%%%%%%%%%%%%%%%%%%%%%%%%%%%%%%%%%%%%%%%

Microgravity experiments were conducted on board the ISS to investigate the long-term flocculation behavior of clay particles in saline water.
For this purpose, ten cuvettes were prepared on Earth and loaded with varying material compositions (cf. \textit{Applied materials}). 
Each cuvette was also equipped with a magnetic bead for stirring purposes to homogenize the suspension.
Before transport to the ISS, the cuvettes were installed in the Binary Colloidal Alloy Test (BCAT) apparatus, a rack in which the cuvettes are arranged in two rows.
An image of the BCAT system is presented in $\S 1$ of the Supplementary Information (SI).
After arrival at the ISS, the system was installed in the Japanese Pressurized Module (JPM), which is part of the Japanese Experiment Module (JEM).
The initial stirring was performed by the astronaut applying a second external magnet and swirling the bead inside the suspension for 60 seconds for each cuvette.
Preliminary tests in an Earth-bound environment showed that this strategy is useful for preparing samples on the ISS \cite{2022_Rommelfanger_etal}. 
After initial homogenization, the cuvettes were left undisturbed for a period of $99$ days and monitored throughout the experiment.

%%%%%%%%%%%%%%%%%%%%%%%%%%%%%%%%%%%%%%%%
\subsection*{Applied materials}
%%%%%%%%%%%%%%%%%%%%%%%%%%%%%%%%%%%%%%%%

Each cuvette was filled with a specified clay composition and salt water, where the latter was the same mixture of purified sodium chloride and deionized water for all cuvettes.
The mixture had a salt concentration of $35 \, [PSU]$ that represents the salinity of seawater \cite{2022_Krahl_etal}, where PSU is the practical salinity unit equivalent to parts per thousand (ppt).
The suspensions contained varying concentrations of kaolinite, montmorillonite, and sand to represent different types of sediment composition \cite{2019_Dohrmann, 2022_Rommelfanger_etal}.
Based on our preliminary earth-bound studies \cite{2022_Rommelfanger_etal,2022_Krahl_etal}, we decided not to vary the salt concentration because its influence is evident only in a very narrow salt concentration range. 
Hence, we decided to go for a high salinity to maximize cohesion and suppress the electric double layer force that can potentially have a repulsive effect.
The specifications of the different compositions are summarized in Table~\ref{tab:runTableExperiments} and further details on the applied materials are specified in $\S 1$ of the SI.
\begin{table}[h!]
  \centering
  \caption{Concentrations of the sediment compositions in $[ppt]$. The compositions in bold represent the samples used for the subsequent analyses, namely cuvettes no. $2$ and $7$.}
    \label{tab:runTableExperiments}
    \begin{tabular}{c | r r r r r }
        \toprule
        Cuvette No. & 1 & {\fontseries{bx}\selectfont 2} & 3 & 4 & 5 \\ 
        \midrule
        Kaolinite & 4 & {\fontseries{bx}\selectfont 8} & 16 & 30 & 12 \\
        Montmorillonite & 12 & {\fontseries{bx}\selectfont 0} & 0 & 0 & 0 \\
        Sand & 60 & {\fontseries{bx}\selectfont 0} & 0 & 0 & 60 \\
        \bottomrule
    \end{tabular}
    \begin{tabular}{c | r r r r r }
        \toprule
        Cuvette No. & 6 & {\fontseries{bx}\selectfont 7} & 8 & 9 & 10 \\ 
        \midrule
        Kaolinite & 0 & {\fontseries{bx}\selectfont 4} & 4 & 4 & 0 \\
        Montmorillonite & 30 & {\fontseries{bx}\selectfont 4} & 12 & 36 & 12 \\
        Sand & 0 & {\fontseries{bx}\selectfont 0} & 0 & 0 & 60 \\
        \bottomrule
    \end{tabular}
\end{table}

%%%%%%%%%%%%%%%%%%%%%%%%%%%%%%%%%%%%%%%%
\subsection*{Image recording and analysis}
%%%%%%%%%%%%%%%%%%%%%%%%%%%%%%%%%%%%%%%%

The samples were continuously recorded by photographs for the entire duration of the experiment.
A Nikon D2Xs camera was installed in front of the rack together with a SB-800 flash on the rear side.
The camera was equipped with the Nikon AF Micro-Nikkor 105/2.8 D lens, which was used in manual focus mode.
Remote shutter release was applied for the recordings to avoid disturbances caused by manually pressing the shutter release.
The interval of image recording was adjusted during the experiments.
Initially, images were taken every $30$ minutes for the first week, then every $2$ hours for the next two weeks, before switching to every $4$~hours for the remainder of the experiments.
The images were analyzed by performing an autocorrelation function to obtain the evolution of the mean floc size \cite{2007_Bailey_etal, 2010_Lu_etal}.
Based on the gray scale of the pixel intensity of the 2D images, a one-dimensional radial autocorrelation is computed to quantify the aggregate sizes over the course of the experiment.
More details on image analysis and applied autocorrelation functions are described in $\S 2$ of SI.

%%%%%%%%%%%%%%%%%%%%%%%%%%%%%%%%%%%%%%%%
\subsection*{Acceleration measurements}
%%%%%%%%%%%%%%%%%%%%%%%%%%%%%%%%%%%%%%%%

\begin{figure}
  \centering
  \includegraphics[width=\linewidth]{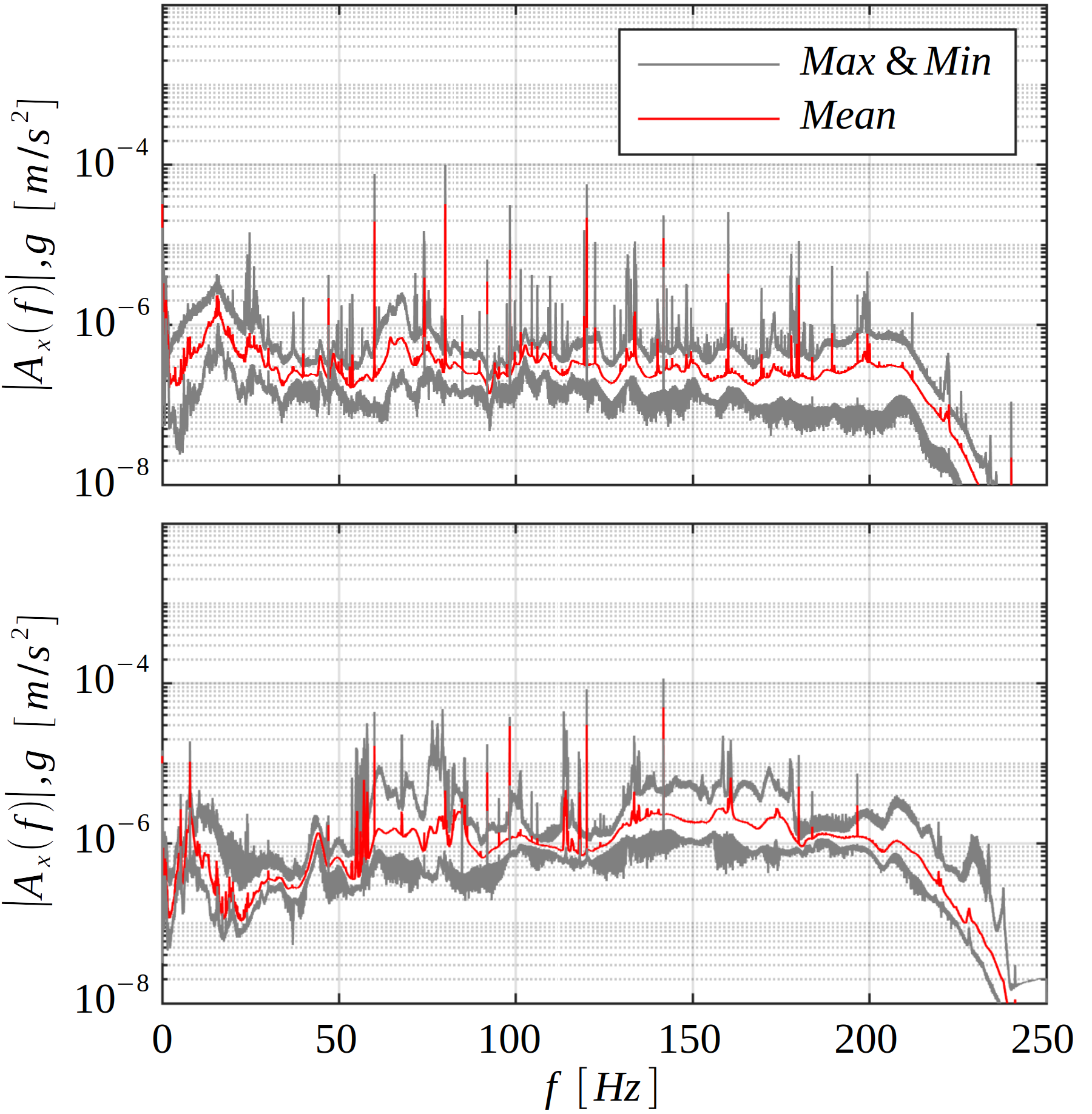}
  \put(-220,230){a)}
  \put(-220,118){b)}
  \caption{Spectra plots of the acceleration and frequency ranges of the g-jitter in $x$-direction, which is parallel to the center line of the laboratory module. The acceleration is expressed as a fraction of the gravitational acceleration $g$. a) presents the data of accelerometer SAMS~$121$f$02$ and b) of SAMS~$121$f$05$.}
  \label{fig:spectraPlots}
\end{figure}

The effects of g-jitter on the experiments were monitored by two accelerometers of the Space Accelerometer Measurement System (SAMS), which were located together with the setup in the JEM of the ISS for the entire duration of the experiment.
The recorded data was made available by the Glenn Research Center(GRC) of the National Aeronautics Space Administration (NASA), based on which the frequency ranges were determined.
Fig. \ref{fig:spectraPlots} presents the analyzed data of the g-jitter of the sensors SAMS~$121$f$02$ (Fig.~\ref{fig:spectraPlots}a) and SAMS~$121$f$05$ (Fig.~\ref{fig:spectraPlots}b) along the $x$-direction.
This direction is parallel to the center line of the laboratory module with its origin in the center of mass of the spacecraft and pointing to the forward direction \cite{2007_Jacobson}.
The spectral plots show the minimum, maximum, and average accelerations expressed as a fraction of the gravitational acceleration $\left(g = 9.81 \, m/s^2 \right)$ for a frequency range between $0$ and $250$~Hz, with local peaks indicating dominant frequencies.
A match of the local peaks in both accelerometers indicates the presence of a decisive source that may have significantly affected the experiments.
This is the case, for example, at around $60$~Hz.
The complementary plots of the remaining coordinate directions, i.e. the $y$-~and~$z$-direction, are presented in $\S 1$ of SI.

%%%%%%%%%%%%%%%%%%%%%%%%%%%%%%%%%%%%%%%%
\subsection*{Numerical simulations}
%%%%%%%%%%%%%%%%%%%%%%%%%%%%%%%%%%%%%%%%

Numerical simulations were performed as an accompanying method to the experiments to investigate the aggregation behavior of cohesive particles in microgravity affected by g-jitter.
The conditions on board the ISS were reproduced on the basis of the analyzed parameter ranges of the experiments.
In this context, cohesive particles were immersed in a cubic container of size $L_{x,y,z} = 20 d_p$, where $d_p$ represents the diameter of an individual particle.
The container was subjected to oscillations to systematically investigate their effects on aggregation behavior within the framework of  pr-DNS \cite{2017_Biegert_etal,2019_Vowinckel_etal,2024_Kleischmann_etal}.
For this purpose, the computational boundaries were characterized by periodic conditions in all directions to exclude influences of the domain walls on particle motion.
We applied the no-slip condition at the particle surfaces.
High spatial and temporal resolutions were achieved by discretizing each particle with $d_p / h = 20$ and each oscillation with $\Delta t = T_f / 200$, where $h$ represents the cell size of the uniform grid, $\Delta t$ the time step, $T_f = 1 / f$ the oscillation period, and $f$ the frequency.

The initial arrangement of the particles ensured a non-contact random distribution.
A cohesive force model was applied to the particles \cite{2019_Vowinckel_etal} and the presence of the g-jitter was modeled by a simplified one-dimensional monochromatic oscillation.
The oscillation direction was defined in the $x$-direction and characterized by $u_f =- A \Omega\sin{\Omega t}$.
Here, $u_f$ represents the velocity of the oscillatory fluid, $A$ the distance amplitude, $A \Omega$ the velocity amplitude, $t$ denotes time, and $\Omega = 2 \pi f$ is the angular frequency \cite{2024_Kleischmann_etal}.
Details of the governing equations and further information about the applied pr-DNS are elaborated in $\S 3$ of SI.

%%%%%%%%%%%%%%%%%%%%%%%%%%%%%%%%%%%%%%%%
\subsection*{Computational scenario}
%%%%%%%%%%%%%%%%%%%%%%%%%%%%%%%%%%%%%%%%

Despite our efforts to extract as much data as possible from the ISS-experiments, a few of the relevant parameters remained unknown to fully characterize the computational scenario. 
Based on our previous work \cite{2024_Kleischmann_etal}, we made appropriate assumptions or applied modified values for some of these parameters to accelerate the flocculation dynamics. 
This was necessary since it is computationally not feasible to entirely reproduce the \mbox{$99$-day} runtime of the experiments with an oscillation frequency of 60~Hz.

The fluid and particle properties that were known from the ISS-experiments are the fluid density $\rho_f = 1,000 \, kg/m^3$, the kinematic viscosity \mbox{$\nu_f = 1.002 \cdot 10^{-6} \, m^2/s$}, as well as the particle density $\rho_p = 2,600 \, kg/m^3$.
Since the initial particle sizes and shapes in the experiments were unknown, the particles were simplified by monodisperse spheres with a chosen diameter of $d_p=115 \, \mu m$.
The frequency was chosen as $f = 60$~Hz according to the measurements of the accelerometers on board. 
If $f$ is considered along with the measured accelerations, which are on the order of \mbox{$|A_x(f)|,g = \orderof(10^{-5}) \, m / s^2 = \orderof(10^1) \, \mu m / s^2$} (cf. Fig.~\ref{fig:spectraPlots}), the corresponding oscillation amplitude is computed by \mbox{$A =|A_x(f)|,g \, / \, \Omega^2 = \orderof(10^{-5}) \, \mu m$}.
Comparing this amplitude to the particle diameter yields an amplitude ratio of \mbox{$\epsilon = A / d_p = \orderof(10^{-7})$}.
In order to increase the particle motion and thus the probability of particle collisions that could lead to faster aggregation, we have increased the oscillation amplitude to \mbox{$A = \orderof(10^{1}) \, \mu m$}.
Furthermore, we have increased the particle volume fraction from $\phi = 0.008$ (see cuvettes no. $2$ and $7$, presented in bold in Table \ref{tab:runTableExperiments}) to an approximate order of magnitude of $\phi = \orderof(10^{-1})$.
An increase in $\phi$ results from an increase in the number of particles $N_p$, which reduces the mean free paths characterized by smaller distances between neighboring particles.
This in turn contributes to the increase in the probability of collisions of particles followed by aggregation.
The exact values of $A$ and $\phi$ for the simulations are provided in the following sections.

%%%%%%%%%%%%%%%%%%%%%%%%%%%%%%%%%%%%%%%%
\subsection*{Non-dimensional quantities}
%%%%%%%%%%%%%%%%%%%%%%%%%%%%%%%%%%%%%%%%

To link our simulations to the ISS experiments, we introduce characteristic scales to non-dimensionalize the numerical properties. 
For this purpose, we choose $d_{p}$ and $1/\Omega$ as relevant length and time scales, as well as $\rho_f$ as density reference.
This yields:

\begin{equation} \label{eq:nonDim_scales}
    \begin{split}
        \begin{gathered}
            \ell = d_{p} \tilde{\ell} \, , \quad 
            t = \tilde{t} / \Omega \,  ,  \quad %t = \frac{\tilde{t}}{\Omega}
            \textbf{u} = d_p \Omega  \tilde{\textbf{u}} \, , \\ %\quad 
            \textbf{D} = d_p^2 \Omega \tilde{\textbf{D}} \, , \quad
            \rho = \rho_f \tilde{\rho}
        \end{gathered}
    \end{split}
\end{equation}
The tilde symbol indicates the dimensionless variables and $\ell$ represents a typical length. 
$\textbf{u}=(u,v,w)^{T}$ is the fluid velocity vector and $\textbf{D}=(D_x,D_y,D_z)^{T}$ the main diagonal of the diffusion tensor.

In addition, we use non-dimensional numbers to generalize the flow properties and to identify dominant physical effects.
Important for the present study are the Reynolds number $Re$, the non-dimensional frequency $S$, and the Stokes number $St$.
The Reynolds number defined as $Re = u_{f,max} d_p / \nu_f$ represents the ratio of inertia and viscous forces, where $u_{f,max} = A \Omega$ is the velocity amplitude.
The non-dimensional frequency describes the characteristics of the oscillations \cite{2001_Coimbra_Rangel,2004_Coimbra_etal,2005_LEsperance_etal,2024_Kleischmann_etal} and we define it as $S = Sl Re = {d_p}^2 \Omega / (36 \nu_f)$, which is a product of the Strouhal number $Sl = \Omega d_p / (9u_f)$ and $Re$.
The Stokes number constitutes the ability of the immersed particles to respond to changes in the fluid flow and is defined as $St = \tau_p / \tau_f = | \rho_s - 1 | \, 2 S$  by comparing the particle response time \mbox{$\tau_p = | \rho_s - 1 | d_p^2 / (18 \nu_f)$} to a characteristic time scale of the fluid flow $\tau_f = 1 / \Omega$~\cite{2024_Kleischmann_etal}.
Here, $\rho_s = \rho_p / \rho_f$ represents the density ratio between the density of the particle and the fluid.

Systematic investigations on the effects of $f$ and $\rho_s$ on the aggregation behavior of particles in oscillations have already been conducted by Kleischmann et al. \cite{2024_Kleischmann_etal}.
Therefore, $S = 0.14$, $St = 0.44$, and $\rho_s = 2.6$ remain constant in all configurations of the present study.
However, the effects of oscillation amplitude and volume fraction on aggregation behavior are largely unexplored. 
The increase in amplitude is expected to increase the mobility of the particles and the increase in volume fraction to reduce the mean free path.
This indicates that both variables might have the potential to significantly enhance aggregate formation.
To investigate this in detail, both parameters were systematically varied within the ranges $\epsilon = [0.05, 0.2]$ and $\phi = [0.042, 0.164]$. 
The variation of $\epsilon$ results in Reynolds numbers spanning the range \mbox{$Re = [0.25, 1]$}.

%%%%%%%%%%%%%%%%%%%%%%%%%%%%%%%%%%%%%%%%
\subsection*{Simulation campaign}
%%%%%%%%%%%%%%%%%%%%%%%%%%%%%%%%%%%%%%%%

The variation of $\epsilon$ and $\phi$ results in different numerical setups, which are shown in Table~\ref{tab:variedNumericalParameters}.
For both parameters, three different values were considered, $\epsilon = 0.05$, $0.1$, and $0.2$, as well as $\phi = 0.042$, $0.084$, and $0.164$.
The latter results in $N_p = 641$, $1283$, and $2502$.
A reference setup (RS) with \mbox{$\phi = 0.084$} and $\epsilon = 0.10$ was defined, to establish a baseline for the comparison of the different parameters.
Based on this, four modified setups were created.  
A setup each with halved (HV) and doubled $\phi$ (DV) while maintaining $\epsilon = 0.10$, as well as a setup each with halved (HA) and doubled $A$ (DA) while keeping $\phi = 0.084$ constant.
\begin{table}[h!]
  \centering
  \caption{Key parameters of numerical simulation scenarios.}
    \label{tab:variedNumericalParameters}
      \renewcommand{\arraystretch}{1.05} 
      \begin{tabular}{c c r c c c }
        \toprule
        ID & $\phi$ & $N_p$ & $\epsilon$  & Scenario \\ \midrule
        RS & $0.084$ & $1283$ & $0.10$ & Reference Setup\\ 
        DV & $0.164$ & $2502$ & $0.10$ & Doubled $\phi$\\ 
        HV & $0.042$ & $641$ & $0.10$ & Halved $\phi$\\ 
        DA & $0.084$ & $1283$ & $0.20$ & Doubled $A$\\ 
        HA & $0.084$ & $1283$ & $0.05$ & Halved $A$\\ 
        \bottomrule
    \end{tabular}
\end{table}

%%%%%%%%%%%%%%%%%%%%%%%%%%%%%%%%%%%%%%%%
%%%%%%%%%%%%%%%%%%%%%%%%%%%%%%%%%%%%%%%%
\section*{Results}\label{sec:results}
%%%%%%%%%%%%%%%%%%%%%%%%%%%%%%%%%%%%%%%%
%%%%%%%%%%%%%%%%%%%%%%%%%%%%%%%%%%%%%%%%

%%%%%%%%%%%%%%%%%%%%%%%%%%%%%%%%%%%%%%%%
\subsection*{ISS Experiments}
%%%%%%%%%%%%%%%%%%%%%%%%%%%%%%%%%%%%%%%%

\begin{figure*}[h!]
  \centering
  \includegraphics[width=0.85\linewidth]{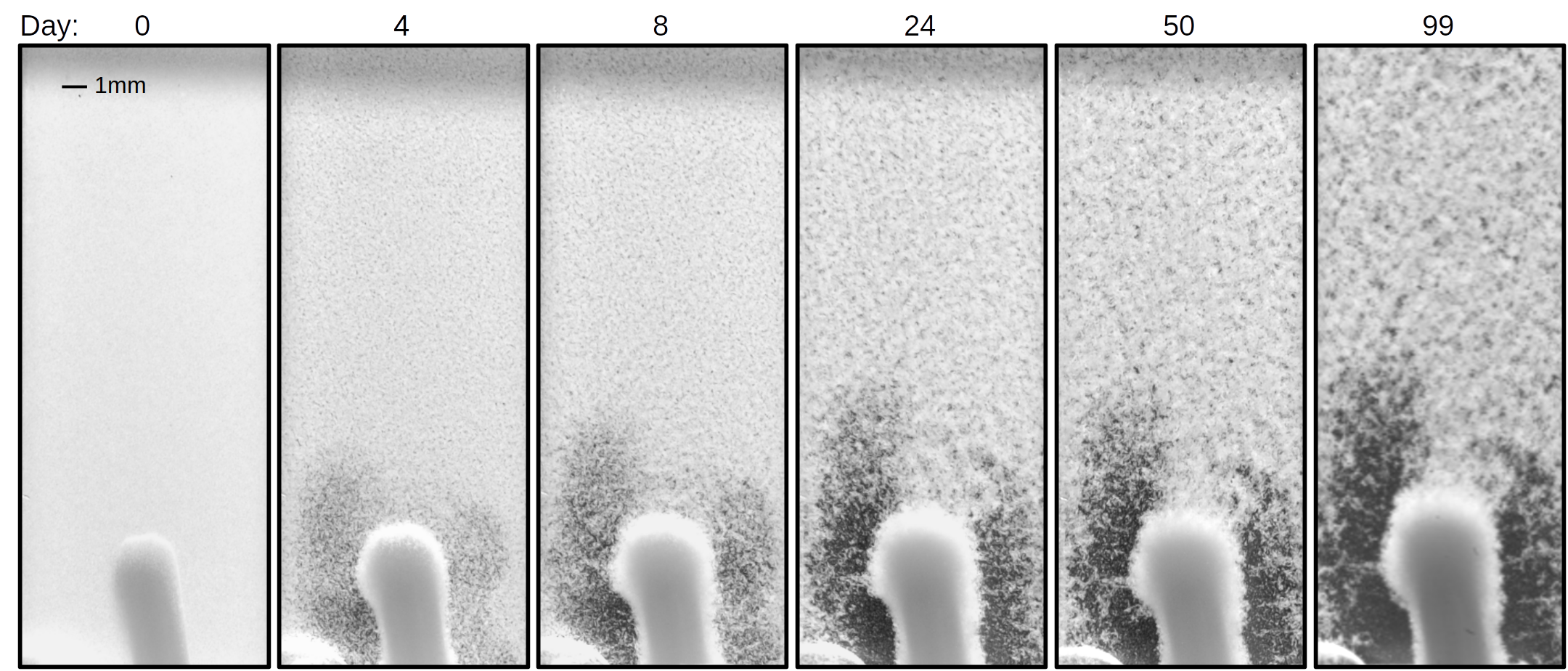}
  \caption{Selected photographs of cuvette no. $2$ over the duration of the experiment with the magnetic bead at the bottom of the cuvette.}
  \label{fig:cuvettes_ISS}
\end{figure*}

Selected photographs of cuvette no. $2$ over the course of the experiments on board the ISS are presented in Fig.~\ref{fig:cuvettes_ISS}. 
Photographs of cuvette no. 7 and detailed close-ups of both cuvettes are shown in $\S 1$ of SI.

A qualitative visual evaluation already reveals that, originating from a homogeneous suspension, pattern formation emerges.
Apparently, the presence of the magnetic bead has a considerable effect on the nearby suspension.
During the course of the experiment, the region close to the magnetic bead is characterized by low particle volume fractions on the left and right edges, while a thin particle strip exists in the center above the magnet.
We hypothesize that this is driven by a flow generated through viscous streaming, induced by the presence of the magnetic bead in the oscillating environment \cite{1955_Lane, 1966_Riley, 1967_Riley}.
Kleischmann et al. \cite{2024_Kleischmann_etal} have shown that such a streaming mechanism can lead to particle aggregation.
Therefore, it can be expected that the viscous flow is present not only around the large magnetic bead, but also around the clay particles far away from it.
Since the focus of this study is not the impact of the magnetic bead on the flocculation, we decided to crop the images so that this effect does not play a role in the analysis we present in the following.
As a result, the image analysis is conducted for a section of the cuvette that is located far above the magnet bead.
In addition, the areas close to the walls as well as the shaded area at the top of the images are excluded from the analysis.

\begin{figure}[h]
  \centering
  \includegraphics[width=\linewidth]{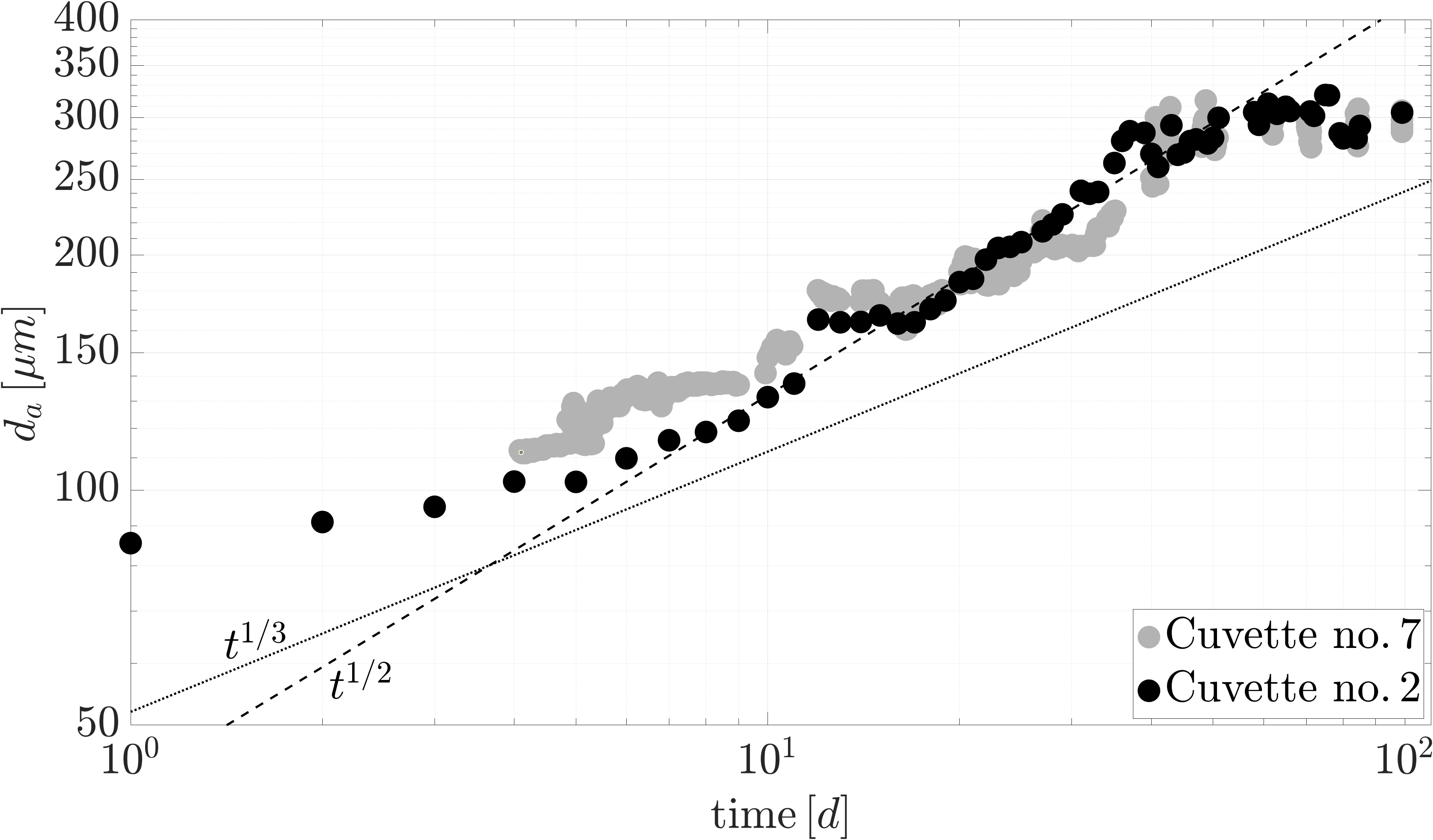}
  \caption{Aggregate growth of cuvettes no. $2$ and $7$. The dashed line represents the scaling $t^{1/2}$ of \eqref{eq:GrwothScaling}. For comparison purposes, the dotted line denotes the scaling $t^{1/3}$ resulting from Brownian motion.}
  \label{fig:aggregateGrwoth_experiments}
\end{figure}

Data processing was performed for cuvettes no. $2$ and $7$ (see in bold in Table~\ref{tab:runTableExperiments}).
These two are the only configurations that satisfy the requirements of a sufficient black and white contrast, while the remaining cuvettes carried too much sediment and appeared largely opaque in the image analysis.
Based on this, a quantitative evolution of the aggregate size over time is calculated for those two cuvettes, as presented in Fig. \ref{fig:aggregateGrwoth_experiments}, where the black and gray circles represent cuvettes no. $2$ and $7$, respectively.
The image conditions in the initial state of cuvette no. $7$ do not allow for a reliable analysis, which is the reason why the results are only shown from day 4 onward.

Both configurations reveal a similar process of aggregate growth that can be divided into an initial phase, steady growth, and a saturation phase.
The transition from the initial phase to the steady growth phase occurs after approximately $7$ days for no.~$2$ and after around $15$ days for no.~$7$.
This transition point of cuvette no. $2$ was selected to apply the aggregate size present at that time as the input parameter for the diameter of the particles in the numerical simulations, which yields $d_p = 115 \, \mu m$.
For both sets of conditions, the steady growth phase lasts until day $50$, after which the aggregate growth slows down and saturates.
The dashed line represents an empirical fit of the growth phase given by
\begin{equation} 
    d_a(t) = \sqrt{\frac{2}{\pi} D  t + {d_{a,0}}^2} \; , 
    \label{eq:GrwothScaling}
\end{equation}
with $d_{a}$ the aggregate size and $d_{a,0}$ the initial aggregate size.
Since the measured oscillations on board the ISS are isotropic (cf. Fig. $2$ in $\S 1$ of SI), it is assumed that the diffusion process is isotropic as well.
This yields the diffusion coefficient $D$ as a scalar that is constant during the steady growth phase, resulting in a time scale of $t^{1/2}$.
Based on the least-squared method, we find \mbox{$D = 3.17 \cdot 10^{-2} \, [\mu m^2/s]$} and $d_{a,0} = 5.43 [\mu m]$ for the empirical fit whose derivation is provided in $\S 4$ of the SI.
The dotted line represents the aggregate growth that can be expected from Brownian motion, for which a size-dependent diffusion coefficient is applied \mbox{$D_B = k_b T / (6 \pi \mu \ell)$} \cite{1905_Einstein}.
Here, $k_B$ is the Boltzmann constant, $T$ the temperature, and $\mu$ the dynamic viscosity.
The size dependence of the diffusion yields a slowdown of the aggregate growth with increasing aggregate size that scales to $t^{1/3}$. Clearly, our data do not follow the scaling due to Brownian motion, suggesting that $D$ is independent of the aggregate size. This suggests that the diffusion is a hydrodynamic effect that may be caused by the g-jitter. This will be discussed in more detail below. 

%%%%%%%%%%%%%%%%%%%%%%%%%%%%%%%%%%%%%%%%
\subsection*{Numerical simulation}
%%%%%%%%%%%%%%%%%%%%%%%%%%%%%%%%%%%%%%%%

In this section, the numerical results of the reference setup RS (cf. Table \ref{tab:variedNumericalParameters}) are presented, before we proceed with the analysis of the setups with modified parameters in the next section.
Note that for all sections of the numerical simulations, the results, equations, and quantities are presented in a dimensionless form according to \eqref{eq:nonDim_scales}.
Therefore, we drop the tilde symbol for the sake of brevity.

\begin{figure}[h]
  \centering
  \includegraphics[width=\linewidth]{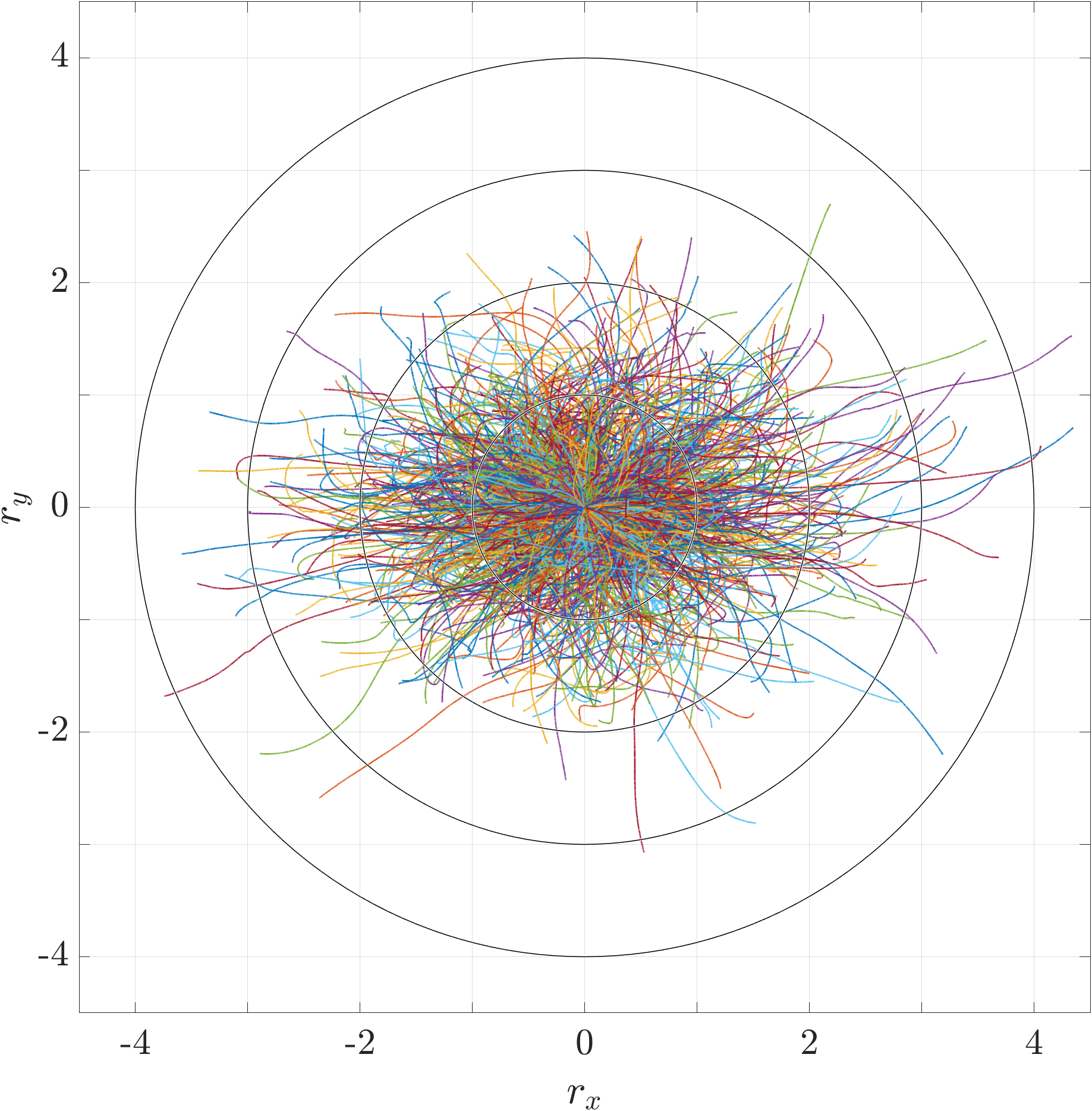}
  \caption{Inertial random walk of RS in a $xy$-plane.}
  \label{fig:randomWalk}
\end{figure}

The wealth of data provided by pr-DNS yields detailed recordings of the individual particle positions at each time step, which allows us to trace and analyze particle trajectories.
To obtain the actual displacement $r_{i,n}$ of each particle $n$ with respect to its initial position $x_{i,n}(0)$ in all coordinate directions $i = (x, y, z)$, we subtract $x_{i,n}(0)$ and the instantaneous mean drift of all particles $\Bar{r}_i$ \cite{2024_Kleischmann_etal} from the actual position of the particle $x_{i,n}$ at a given time $t$:
\begin{equation} 
    r_{i,n}(t) = x_{i,n}(t) - x_{i,n}(0) - \Bar{r}_i(t)
    \label{eq:displacement}
\end{equation}
The total distances traveled by the particles throughout the entire simulation ($t/T_f = 25,000$) are presented in Fig. \ref{fig:randomWalk} for the $xy$-plane, where the concentric circles indicate the distances traveled.
The illustration shows a more pronounced expansion of the particle displacements in the direction of oscillation, namely the $x$-direction, which is about twice as large as in the other directions.
The displacements also reveal the randomness of the particle motion with each particle following a unique path.
Illustrations of the displacements separated for each coordinate direction are shown in $\S 5$ of SI.

\begin{figure}[b!]
  \centering
  \includegraphics[width=\linewidth]{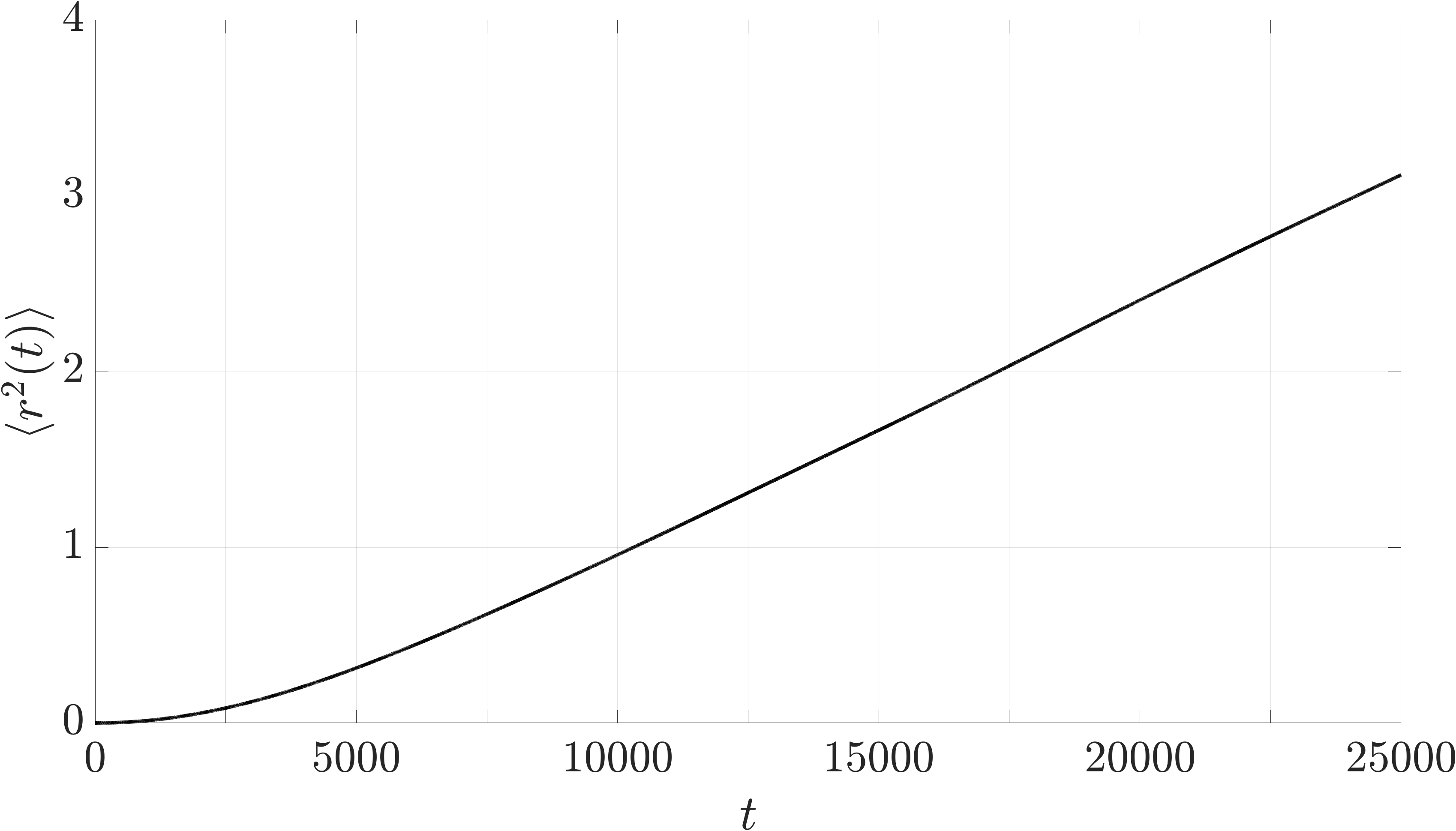}
  \caption{Mean square displacement $\langle r^2(t) \rangle$ of RS.}
  \label{fig:MeanSquareDisplacement}
\end{figure}

We compute the mean particle displacements by the mean square displacement $\langle r^2(t) \rangle$ \cite{1991_Qian_etal} to characterize the dynamics of the particles
\begin{equation} 
    \langle r^2(t) \rangle = \sum_{i=1}^{3} \left[ \frac{1}{N}  \sum_{n=1}^{N} \left| r_{i,n}(t) \right|^2 \right] \, , 
    \label{eq:MSD}
\end{equation}
where $N$ is the total number of particles.
The resulting evolution of $\langle r^2(t) \rangle$ over time is illustrated in Fig.~\ref{fig:MeanSquareDisplacement}.
The figure shows an initial trend characterized by $\langle r^2(t) \rangle \propto t^2$.
This trend is generally classified as ballistic motion, where the particles tend to move without interacting with each other.
As soon as particle-particle collisions occur, the trend transitions to a linear progression ($\langle r^2(t) \rangle \propto t$) which is defined as diffusive motion \cite{2000_Blum_etal, 2011_Huang_etal, 2019_Riahi_etal}.

\begin{comment}
Based on the diffusive motion, we calculate the diffusion coefficient $D$ given by \cite{1905_Einstein}
\BV{And now you are switching again to isotropic diffusion}
%
\begin{equation} 
    \langle r^2(t) \rangle = 2 D t  \; .
    \label{eq:DiffusionCoefficient}
\end{equation}

It is important to note that $\langle r^2(t) \rangle$ is the sum of the individual components of the coordinate directions $j$
%
\begin{equation} \label{eq:MSD_components}
   \langle r^2(t) \rangle = \sum_{j=1}^{3} \langle r^2_j(t) \rangle = \langle r^2_x(t) \rangle + \langle r^2_y(t) \rangle + \langle r^2_z(t) \rangle  \; ,
\end{equation}
%
which also applies to $D$ when focusing on the principal components and thus results in
%
\begin{equation} \label{eq:D_components}
  % 2 D t = \left[\sum_{j=1}^{3} D_j \right] 2 t =(D_x + D_y + D_z ) \; 2t  = D_x + D_y + D_z \; ,
  D_j = \langle r^2_j(t) \rangle / (2t)\; .
\end{equation}
%
Calculating the principal components of the diffusion tensor yields $D_x = 3.61$, $D_y = 1.75$ and \mbox{$D_z = 1.90 \; [\cdot 10^{-5}]$}.
\BV{Here and in the entire document: it should not be the principal components, but the main diagonal components of the diffusion tensor.}

\end{comment}

\begin{figure*}[h]
  \centering
  \includegraphics[width=0.85\linewidth]{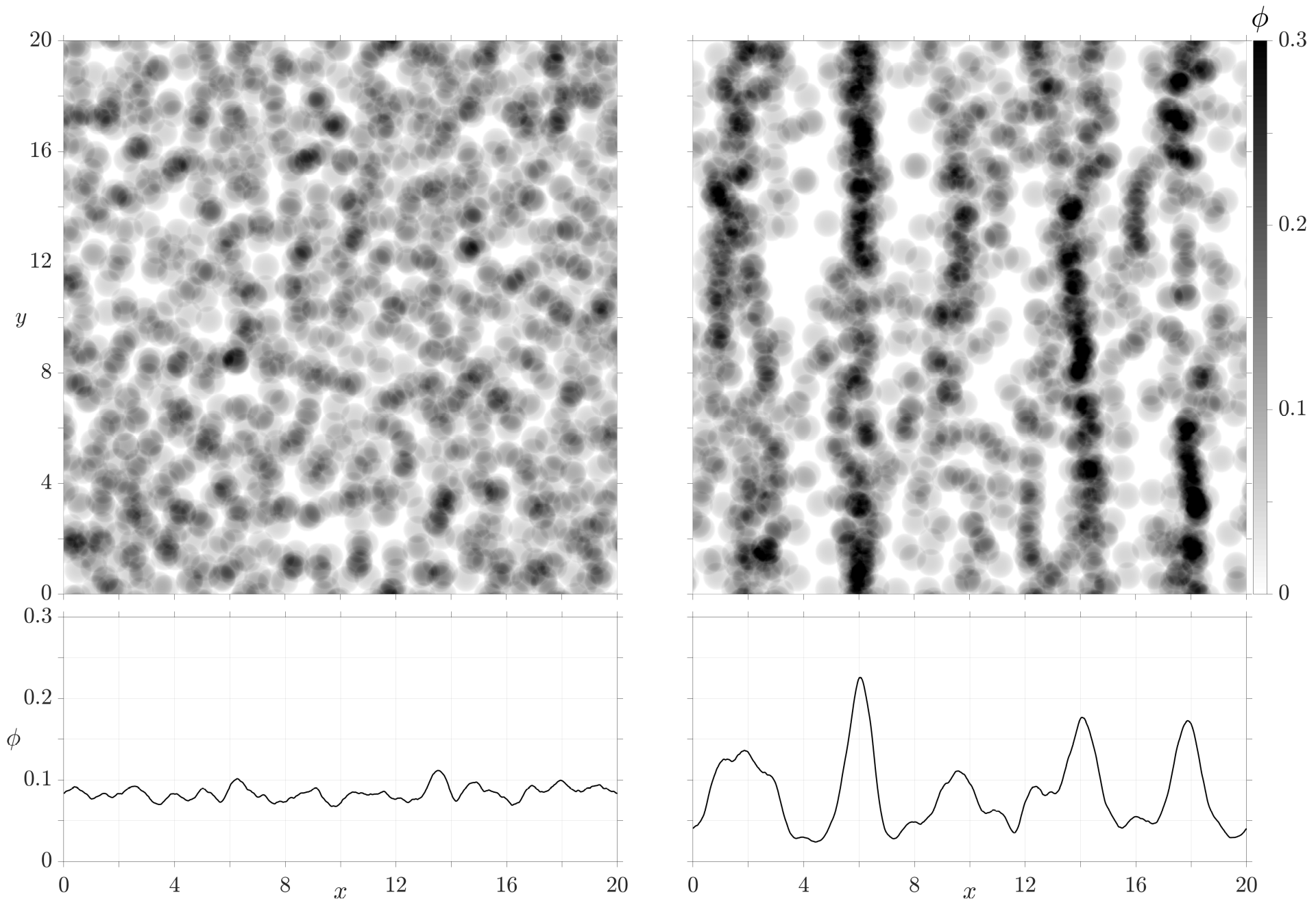}
  \put(-410,260){a)}
  \put(-208,260){b)}
  \caption{Volume fractions of the two-dimensional $xy-$plane (upper) and one-dimensional along the $x-$axis (lower) of RS. a) shows the initial setup at $t = 0$ and b) the final arrangement at $t = 25000$.}
  \label{fig:volumeFractions_OS}
\end{figure*}

In addition to the quantitative and empirical analysis of the particle motion, we also qualitatively analyze the particle behavior.
To this end, we visualize the volume fractions, which provide information on the distributions of the particles and allow the identification of structures of aggregated particles.
Fig.~\ref{fig:volumeFractions_OS} shows the volume fraction averaged in the $z$-direction in a 2D plane in the upper panel and the 1D volume fraction along the $x$-direction averaged in the $y$- and $z$-directions in the lower one.
For the 2D planes, the volume fractions are indicated by the gray scale, ranging from $0$ to $0.3$, where Fig. \ref{fig:volumeFractions_OS}a presents the initial arrangement and Fig. \ref{fig:volumeFractions_OS}b the final state after $25,000$ oscillation periods.
The profile of the 1D volume fraction fluctuates around the value $\phi = 0.084$ in the initial arrangement, accompanied by an almost homogeneous particle arrangement in the 2D representation.
In the final state, on the other hand, we observe the formation of distinct chain-like structures that form perpendicular to the oscillation direction.
Elongated clusters of particles (2D), accompanied by local peaks in the 1D representation, are adjacent to less populated sections, visualized by local troughs.

Next, we examine the aggregate growth generated by the unidirectional oscillations by applying a normalized autocorrelation function ($C$) to the 1D volume fraction.
\begin{equation}
    \label{eq:1D_autocorrelation}
    C(r) = \frac{\sum_i (\phi_i - \Bar{\phi})(\phi_{i+r}-\Bar{\phi})}{\sum_i (\phi_i - \Bar{\phi})^2}
\end{equation}
Here, $r$ is the position shift (lag) along the $x$-axis, $i$~is the position index, and $\Bar{\phi}$ the mean of the $1$D volume fraction.
Applying \eqref{eq:1D_autocorrelation} yields an autocorrelation ranging between \mbox{$-1$ and $1$}.
Starting from $C(0) = 1$, the first local trough of the graph determines the characteristic length scale ($d_a$), which describes the size of an aggregate.
More details and a graphical representation of $C$ can be found in $\S 2$ of the SI.

\begin{figure}[h]
  \centering
  \includegraphics[width=\linewidth]{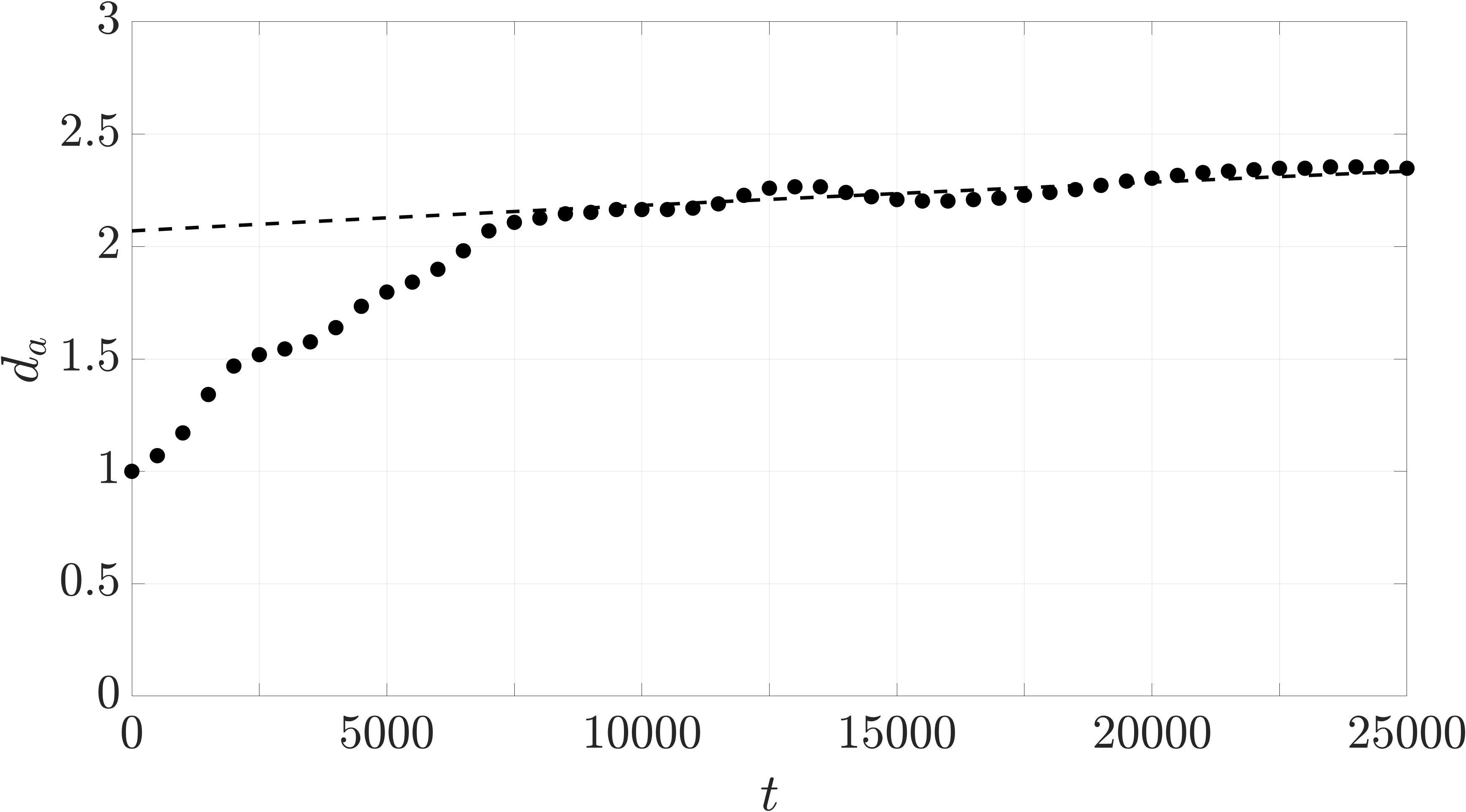}
  \caption{Aggregate size over time. The black circles represent the numerical results and the dashed line the empirical fit \eqref{eq:GrwothScaling}.}
  \label{fig:aggregateGrwoth_numerical}
\end{figure}

The aggregate size $d_a$ is evaluated for every $500^{th}$ oscillation period and is presented in Fig.~\ref{fig:aggregateGrwoth_numerical} by the black circles.
The empirical fit \eqref{eq:GrwothScaling} is presented by the dashed line, representing the expected growth rate that scales by $t^{1/2}$.
Its applicability is substantiated in the next section through a detailed discussion of the different variants.
Here, $d_{a,0}$ is calibrated to achieve a successful agreement between the empirical fit and the numerical results of RS. 
We apply \eqref{eq:MSD} to compute the trace of the main diagonal of the diffusion tensor to obtain the diffusion coefficient $D$ given by \cite{1905_Einstein}
\begin{equation} 
    \langle r^2(t) \rangle = 2 D t  \; .
    \label{eq:DiffusionCoefficient}
\end{equation}
In order to transfer the isotropic conditions of the experiments to the numerical simulations, we assume that the contributions of $D_y$ and $D_z$ would be equivalent to $D_x$ in a 3D oscillation.
Therefore, we simplify as follows:
\begin{equation} \label{eq:D_eff}
  D' =D_x + D_y + D_z \approx 3 D_x \; .
\end{equation}
Applying this assumption, we denote the diffusion coefficient by $D'$ to clarify the distinction from the resulting diffusion coefficient of the experiments.

The trend of numerical aggregate growth presented in Fig.~\ref{fig:aggregateGrwoth_numerical} can be divided into different growth phases, similar to the results of the ISS experiments.
Starting with an initial phase, where aggregates begin to form rapidly, the growth rate progresses to a phase of constant growth.

%%%%%%%%%%%%%%%%%%%%%%%%%%%%%%%%%%%%%%%%
\subsection*{Numerical variants}
%%%%%%%%%%%%%%%%%%%%%%%%%%%%%%%%%%%%%%%%

The numerical variants provide an opportunity to investigate the impact of the modifications of $\epsilon$ and $\phi$ on the aggregation mechanism.
In this regard, the same analysis methods that generated the results of RS were applied.
In detail, we focus on the comparison of $\langle r^2(t) \rangle$ and the growth rates of the aggregates together with the respective diffusion coefficients in this section.

\begin{figure*}[h!]
  \centering
  \includegraphics[trim=1cm 1cm 0.4cm 1.3cm,clip,width=\textwidth]{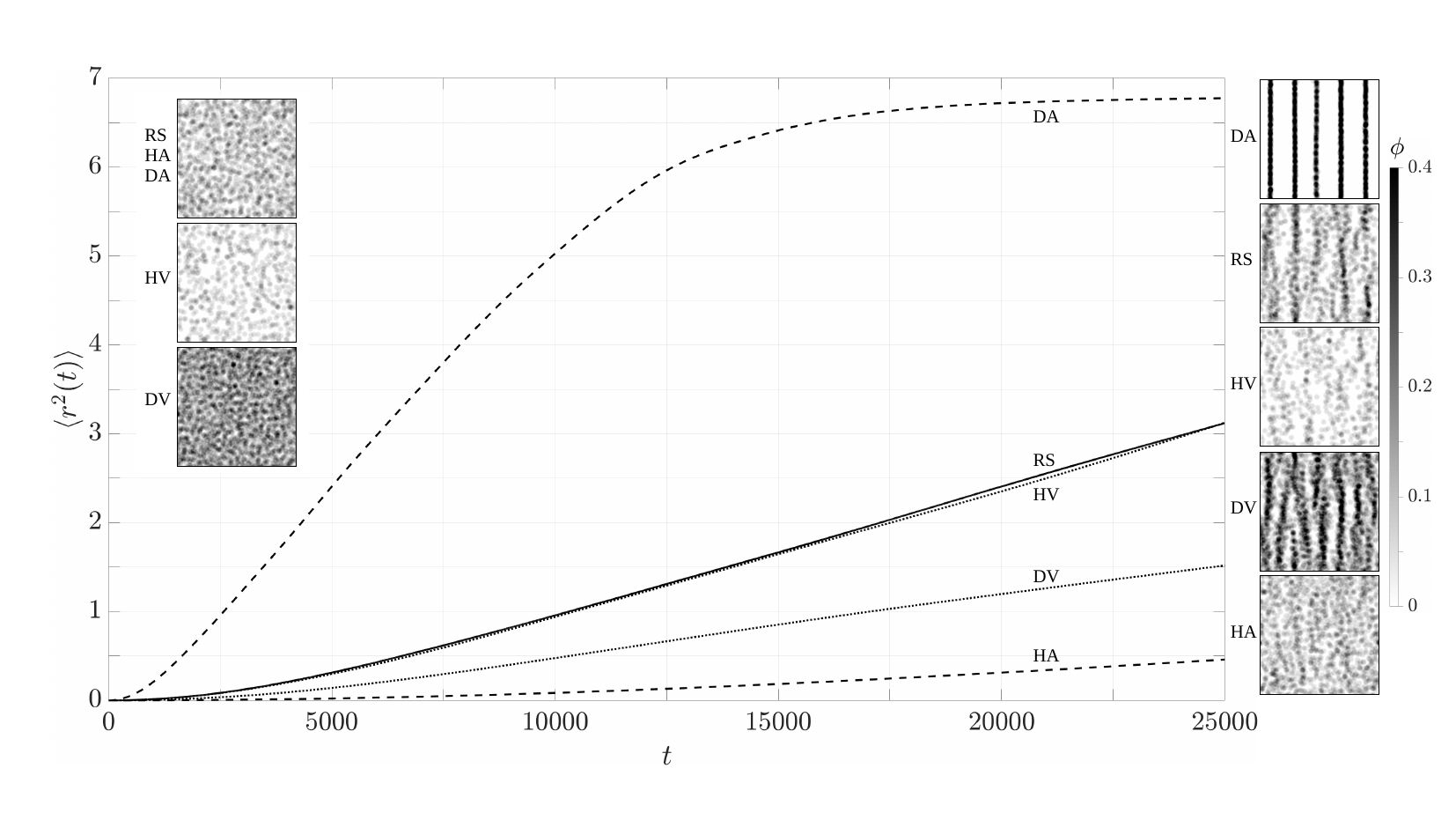}
  \caption{Comparison of the mean square displacements $\langle r^2(t) \rangle$ of the numerical variants over time. The subfigures illustrate the 2D volume fractions $\phi$ of the respective setups in black and white according to the scale on the right. The left panel presents the initial setups, where RS, HA and DA have the same configuration, and the right panel depicts the end of the simulations.}
  \label{fig:MSD_comparison}
\end{figure*}

Fig. \ref{fig:MSD_comparison} shows $\langle r^2(t) \rangle$ over time, where RS is illustrated by the solid line (cf. Fig. \ref{fig:MeanSquareDisplacement}), the cases with modified amplitudes HA and DA are shown as the dashed lines, and the cases with the modified volume fraction HV and DV are represented by the dotted lines.
The accompanying diffusion coefficients are calculated based on \eqref{eq:DiffusionCoefficient} and \eqref{eq:D_eff} and are presented in Table~\ref{tab:Results_diffusionCoefficients}.
The trend of the modified volume fractions is similar to RS with almost the same progression of HV and a flatter curve of DV, where the slope of this curve is approximately half of those of RS and HV.
This is reflected in the results of $D'$, where the outcome of HV ($D'=10.72 \cdot 10^{-5}$) is similar to RS ($D'=10.84 \cdot 10^{-5}$), while DV ($D'=5.32 \cdot 10^{-5}$) is almost half.
In contrast, the modified amplitudes show clear differences compared to RS.
DA shows a significantly higher increase at the beginning of the simulations, which is reflected in a higher value of $D'=45.33 \cdot 10^{-5}$.
After around $12,500$ oscillations,  $\langle r^2(t) \rangle$ starts to stagnate for DA and converges to approximately $\langle r^2(t) \rangle = 6.75$.
On the other hand, HA shows a lower slope and is significantly below RS with a smaller diffusion \mbox{coefficient ($D'=2.06 \cdot 10^{-5}$)}.
\begin{table}[h!]
  \centering
  \caption{Diffusion coefficients of the numerical variants based on the assumption in~\eqref{eq:D_eff} in the scale of $10^{-5}$.}
    \label{tab:Results_diffusionCoefficients}
      \renewcommand{\arraystretch}{1.05} 
      \begin{tabular}{l | c c c c c}
        \toprule
        ID & RS & DV & HV  & DA & HA \\ \midrule
        $D'$ & $10.84$ & $5.32$ & $10.72$ & $45.33$ & $2.06$\\
        \bottomrule
    \end{tabular}
\end{table}

\begin{figure}
  \centering
  \includegraphics[width=\linewidth]{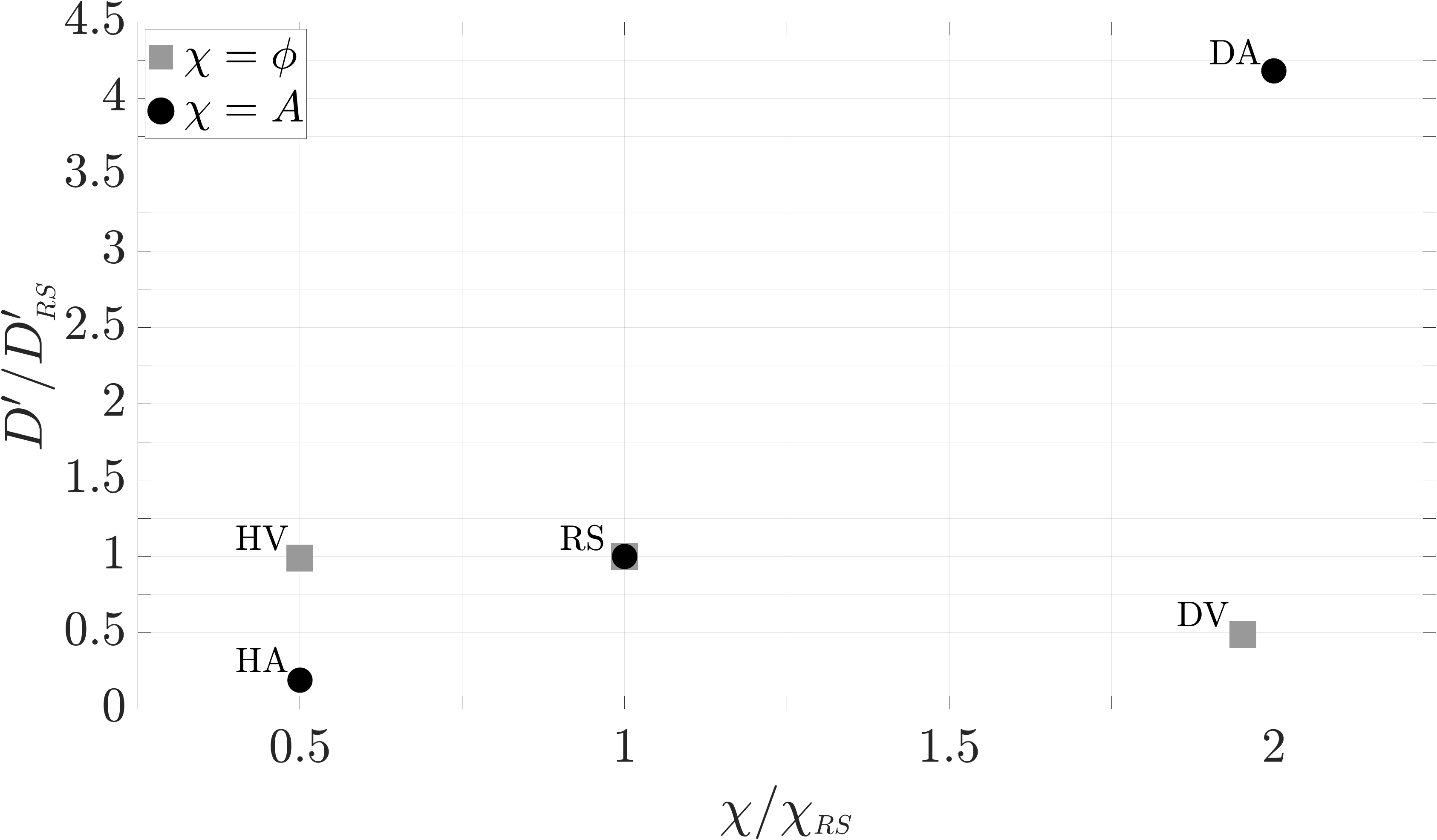}
  \caption{Representation of $D' / D'_{RS}$ against $\chi / \chi_{RS}$, where $\chi$ denotes the parameter of the respective variant, resulting in $\phi / \phi_{RS}$ or $A / A_{RS}$. }
  \label{fig:D_Coeffs_ratio}
\end{figure}

Of particular importance here is the ratio of $D'$ between each variant and RS ($D'_{RS}$).
Figure \ref{fig:D_Coeffs_ratio} presents the ratios of the diffusion coefficients $D' / D'_{RS}$ for the different variants. 
These ratios are plotted against $\chi / \chi_{RS}$, which represents the ratio of the variable parameter to RS ($\phi / \phi_{RS}$ or $A / A_{RS}$), depending on the variant.
The presentation of RS is for the purpose of completeness unity.
The result of HV is also approximately~$1$, while the ratio of DV is close to $0.5$.
The diffusion coefficient of HA is almost five times smaller than RS, yielding a ratio of $0.19$.
In contrast, DA is more than four times larger with a ratio of $4.18$.

The mean square displacements of the particle motion (Fig. \ref{fig:MSD_comparison}) show a similar behavior for all numerical variants despite obvious differences in the magnitudes of $\langle r^2(t) \rangle$ and the results of $D$.
Each trend starts with an initial ballistic phase ($\langle r^2(t) \rangle \propto t^2$) before transitioning to a linear diffusive phase ($\langle r^2(t) \rangle \propto t$).
In the case of DA, this linear progression ends even during the course of the simulation and turns into stagnation.
The other variants, which are RS, HA, HV, and DV, are still in the diffusive phase at the end of the simulations.

The subfigures of Fig. \ref{fig:MSD_comparison} illustrate the 2D volume fractions with the initial configurations on the left and the results after $25,000$ oscillations on the right. 
The tags indicate the affiliation of the configurations with the respective setup, where RS, HA, and DA have the same initial particle arrangement.
The resulting configurations are ordered according to the results of $\langle r^2(t) \rangle$ and the gray scale is represented by the black and white scheme on the right.
Note that the illustrations of RS correspond to Fig. \ref{fig:volumeFractions_OS}, but are shown again for comparison purposes. 
The resulting volume fractions of the variants highlight the impacts of the modified parameters.
Fully developed chain structures are only present in DA, where distinct gaps without particles are between the stripes.
The setups of RS and DV show the first formations of chains, in which, however, many particles are still not assigned to any chain and are therefore located in the intermediate spaces.
No obvious changes in the final particle arrangement are apparent for HA and HV compared to their initial condition.

\begin{figure}[h]
  \centering
  \includegraphics[width=\linewidth]{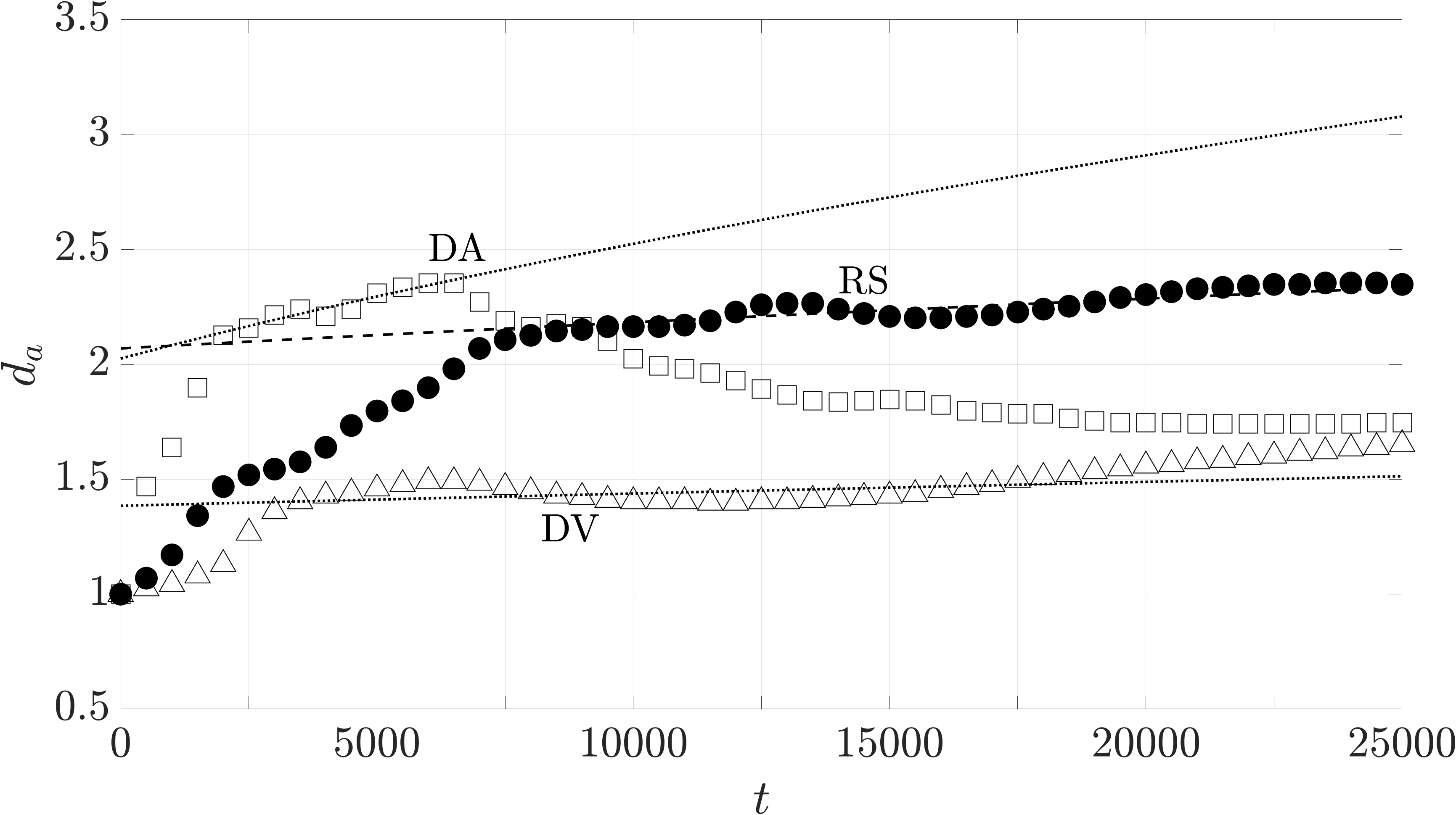}
  \caption{Comparisons of the aggregate growths with the respective empirical fits for the setups RS, DA, and DV. }
  \label{fig:aggregateGrowth_comparison}
\end{figure}

Based on the 1D volume fractions shown for all variants in $\S 6$ of SI, $d_a$ is calculated by applying \eqref{eq:1D_autocorrelation}.
However, since recognizable structures are required for this analysis, only the setups RS, DA, and DV are considered.
The aggregate sizes over time, together with the respective empirical fits, are shown in Fig. \ref{fig:aggregateGrowth_comparison}.
The results of RS are the same as in Fig. \ref{fig:aggregateGrwoth_numerical} but are presented again for the sake of comparison.
The results of the variants presented are tagged by the respective ID, where the aggregate sizes are given by the symbols, and the empirical fits are illustrated by the lines.
For the latter, the diffusion coefficients of Tab. \ref{tab:Results_diffusionCoefficients} are applied to \eqref{eq:GrwothScaling} and calibrations are performed for $d_a$ of the respective variants.
The development of the aggregate size of DV shows the same characteristics as in RS, with an initial phase of rapid aggregate growth and a subsequent progression that shows a linear trend despite slight fluctuations.
The aggregate growth of DA represents three phases that have already been identified for the trend of $\langle r^2(t) \rangle$ in Fig.~\ref{fig:MSD_comparison}.
The process of aggregate growth starts with a short period of rapid increase, followed by a growth phase consistent with the associated empirical fit, before it decreases and reaches the saturation phase.

%%%%%%%%%%%%%%%%%%%%%%%%%%%%%%%%%%%%%%%%
%%%%%%%%%%%%%%%%%%%%%%%%%%%%%%%%%%%%%%%%
\section*{Discussion}\label{sec:discussion}
%%%%%%%%%%%%%%%%%%%%%%%%%%%%%%%%%%%%%%%%
%%%%%%%%%%%%%%%%%%%%%%%%%%%%%%%%%%%%%%%%

The investigation of cohesive sediments immersed in salt water and subjected to \mbox{g-jitter} in microgravity shows that particle aggregation can occur solely due to the effect of oscillations.
The design of the experiments on board the ISS enabled the exclusion of the aggregation mechanisms of shear flow and differential settling that are otherwise prevalent on Earth.
Shear flow was eliminated by utilizing closed cuvettes and differential settling as a result of the microgravity environment, where particles and aggregates do not settle.
The same conditions were established in the numerical simulations that were performed on the basis of pr-DNS.
The evaluation of aggregate growth in both physical experiments and numerical simulations reveal a growth rate that scales with $t^{1/2}$ (cf. Figs. \ref{fig:aggregateGrwoth_experiments}, \ref{fig:aggregateGrwoth_numerical}, and \ref{fig:aggregateGrowth_comparison}).
The application of the empirical fit \eqref{eq:GrwothScaling} showed that the diffusion coefficients $D$ and $D'$, respectively, were constant in both settings after an initial ballistic phase.
This shows that the impact of oscillations exceeds the potential effect of Brownian motion in the ISS-experiments that would have been characterized by a growth rate that scales with $t^{1/3}$ and a diffusion coefficient that varies over time depending on the actual size of the aggregates \cite{1905_Einstein}.
In this regard, it can be concluded that in addition to differential settling and liquid shear, Brownian motion is also not of significance for the aggregation process in the investigations presented.

In order to understand aggregation through g-jitter vibrations, we first need to examine the effect of oscillations on each individual particle.
Therefore, we analyzed detailed recordings of the particle trajectories from the numerical simulations (cf. Fig. \ref{fig:randomWalk}).
The data showed that the impact of oscillations leads to a unique pattern of motion of each particle, known as an inertial random walk \cite{1996_Trolinger_etal}.
The randomness of their trajectories arises from the presence of neighboring particles, where the inertia of each individual particle causes its trajectory to deviate from the fluid flow, altering the surrounding flow field and affecting the behavior of adjacent particles \cite{2013_Lappa}.
In a dense system, the flow fields generated by the particles interact with each other, causing each particle to experience a unique flow condition and to promote collisions that might lead to aggregation.
As soon as particles collide, they remain together due to cohesion and form larger aggregates.
As aggregates grow in size, their inertia increases, causing greater deviations from the fluid flow, resulting in even larger flow structures around the respective aggregate.
This expands the radius of influence on neighboring particles, increasing the likelihood of collisions and promoting further growth.

The development of aggregate growth can be divided into three phases: an initial phase, steady growth, and a saturation phase.
This categorization is identical in the experimental and numerical results, despite differences within the individual phases.
For example, the tendency for aggregate growth differs between the experimental and numerical results in the initial phase.
In the experiments, the particle dynamics in the initial phase is slower than in the steady growth phase, where we obtain the proportionality to $t^{1/2}$.
It can be assumed that the particles are more dispersed at the beginning of the experiment, i.e. after homogenization of the suspension, and therefore the particle collisions that are essential for aggregate growth do not occur as frequently.
Once collisions occur and aggregates form, these aggregates generate larger flow structures that have a larger volume of influence, which in turn increases the rate of aggregation.
In contrast, the numerical results show the opposite behavior, in which the initial aggregation is characterized by rapid growth before it takes on a linear progression of the steady growth phase.
In this context, it is important to note that the numerical realization of the oscillations differs from the conditions on board the ISS.
In the numerical simulations, the oscillations are unidirectional and monophasic, whereas the g-jitter on the ISS is multidirectional and more random.
Hence, for the simulations we obtain that the one-dimensional oscillation leads to a pronounced particle movement along the direction of oscillation, while the effect of dispersion along the other two axes is limited. 
This behavior can be recognized in the trajectories of the particles in Fig.~\ref{fig:randomWalk}.

As soon as aggregates are formed, two aspects become relevant that could be responsible for the reduction of the ongoing aggregation process.
First, particle clusters might experience reduced accelerations due to their increased size that slow down the collision frequency.
Second, the aggregates tend to align perpendicular to the oscillation direction, reducing their exposure to incoming particles and thus the probability of further collisions.
The latter was shown by visualization of the particle volume fractions of the numerical simulations in Figs. \ref{fig:volumeFractions_OS} and \ref{fig:MSD_comparison}, where the chain formations were already visible or even fully developed (e.g., in the setup DA) at the end of the simulation time.
The formation of particle chains oriented perpendicular to the oscillation direction was already shown by previous experimental and numerical studies that investigated non-cohesive spherical particles subjected to 1D oscillations on a frictionless plate under earth-bound conditions \cite{2009_Klotsa_etal, 2016_Mazzuoli_etal, 2023_vanOverveld_etal}.
These studies also revealed that the particles tend to remain within the chain structure and generally do not move between the respective chains.
Van Overveld and colleagues \cite{2023_vanOverveld_etal, 2024_vanOverveld_etal} were able to show that this is due to hydrodynamic forces that lead to long-range attraction and short-range repulsion, causing chains to keep a certain distance from each other.
This effect also explains the clear chain formation for the setup DA.
As soon as particles belong to one chain, they move together and do not exchange with neighboring structures, causing a convergence of $\langle r^2(t) \rangle$ to a constant value after $t/T_f=12,500$ (Fig. \ref{fig:MSD_comparison}).
After the chains have formed, the particles align themselves in such a way that the particles of a chain are as close to a line as possible.
This in turn is recognized as a decrease in aggregate size after previous growth (Fig.~\ref{fig:aggregateGrowth_comparison}), when determining aggregate size by applying the autocorrelation function \eqref{eq:1D_autocorrelation} to the one-dimensional volume fractions.
In the numerical setups RS and DV, the chain structures are only slightly recognizable if at all, and many particles are still widely dispersed. 
Therefore, $\langle r^2(t) \rangle$ and $d_a$ are still increasing and have not yet converged to a constant value.

An examination of the two modified parameters $\phi$ and $A$ reveals that both have an impact on the particle kinematics in the numerical simulations.
The volume fraction of the particles is accompanied by the number of particles and determines the mean free path characterized by the average length available for an individual particle to travel without interference from surrounding particles.
Consequently, an increase in $\phi$ decreases the mean free path and, vice versa, a decrease in $\phi$ leads to a larger path.
This is consistent with $\langle r^2(t)\rangle$ of DV, the configuration with a doubled volume fraction, where the displacement of the particles decreases due to the increase in the number of particles (cf. Fig. \ref{fig:MSD_comparison}).
This in turn results in slower aggregate growth (Fig. \ref{fig:aggregateGrowth_comparison}).
The setup HV with halved $\phi$ leads to a trend of $\langle r^2(t)\rangle$ similar to  the reference setup (RS), although it would have been expected that the displacement would be greater due to a larger mean free path.
The available data do not provide conclusions about this behavior, but a hypothesis could be that the distances between the particles are of a size at which the existing flow conditions cause only minor interactions between the particles. 
In fact, our preliminary study \cite{2024_Kleischmann_etal} confirms that there is an exponential decay of particle interaction due to viscous streaming induced by high-frequency oscillations. 
This could also explain the fact that almost no aggregates form within the simulation time for HV, which is evident from the very similar initial and final configurations of the 2D volume fractions (cf. subfigures of Fig. \ref{fig:MSD_comparison}).
These findings suggest that there could be an optimal volume fraction for aggregate growth.
Our results indicate that this critical volume fraction could possibly be close to the value of $\phi$ of the setup RS.

The second parameter is the oscillation amplitude $A$, which is decisive for the excursion of the oscillation and consequently for the motion of the particles.
Depending on the particle excursion, the probability of particle-particle interactions either increases or decreases.
This is because the flow structure range around an individual particle increases depending on the excursion distance.
The results of the halved and doubled amplitudes, HA and DA, reflect this behavior.
HA has a significantly reduced $\langle r^2(t)\rangle$ compared to RS, while DA shows a much more pronounced displacement of the particles (Fig. \ref{fig:MSD_comparison}).
The smaller particle displacements of HA lead to fewer particle-particle interactions, which reduces the probability of aggregate formation.
Similarly to HV, this does not lead to apparent changes in the 2D volume fractions that are presented in the subfigures of Fig. \ref{fig:MSD_comparison}.
The increased amplitude of DA increases the excursion of the particles leading to a higher probability of particle collisions, resulting in significantly more pronounced and faster aggregate formation (Figs. \ref{fig:MSD_comparison} and \ref{fig:aggregateGrowth_comparison}).
This reveals that the amplitude has a considerable influence on the behavior of the particles and consequently on the aggregate growth.
The results presented suggest that increasing the amplitude may enhance aggregate formation. 
However, it is important to consider that aggregates can also break apart when hydrodynamic forces exceed cohesive binding forces \cite{2020_Halfi_etal, 2025_Bendory_Friedler}.

We can conclude that oscillations have a significant effect on particles immersed in a fluid and that they are promising in accelerating and controlling the aggregation of suspensions.
The investigated parameters of the volume fraction and the oscillation amplitude demonstrate that several factors have an effect on the aggregation process.
Therefore, there is the potential to perform precise analyses in future studies, on the basis of which the optimum conditions for aggregate formation could be derived.
Such conditions would significantly improve this method, making it suitable for a wide range of applications, such as drinking water treatment or industrial processes.

%%%%%%%%%%%%%%%%%%%%%%%%%%%%%%%%%%%%%%%%
%%%%%%%%%%%%%%%%%%%%%%%%%%%%%%%%%%%%%%%%
\section*{Data Availability}
%%%%%%%%%%%%%%%%%%%%%%%%%%%%%%%%%%%%%%%%
%%%%%%%%%%%%%%%%%%%%%%%%%%%%%%%%%%%%%%%%

The datasets generated and analyzed during the current study are available from the corresponding author upon reasonable request.

%%%%%%%%%%%%%%%%%%%%%%%%%%%%%%%%%%%%%%%%
%%%%%%%%%%%%%%%%%%%%%%%%%%%%%%%%%%%%%%%%
\section*{Code Availability}
%%%%%%%%%%%%%%%%%%%%%%%%%%%%%%%%%%%%%%%%
%%%%%%%%%%%%%%%%%%%%%%%%%%%%%%%%%%%%%%%%

The underlying codes and datasets for this study are available from the corresponding author upon reasonable request.

%%%%%%%%%%%%%%%%%%%%%%%%%%%%%%%%%%%%%%%%
%%%%%%%%%%%%%%%%%%%%%%%%%%%%%%%%%%%%%%%%
% References
%%%%%%%%%%%%%%%%%%%%%%%%%%%%%%%%%%%%%%%%
%%%%%%%%%%%%%%%%%%%%%%%%%%%%%%%%%%%%%%%%

%\bibliographystyle{numeric}
%\bibliography{bib}
\printbibliography

%%%%%%%%%%%%%%%%%%%%%%%%%%%%%%%%%%%%%%%%
%%%%%%%%%%%%%%%%%%%%%%%%%%%%%%%%%%%%%%%%
\section*{Acknowledgments}
%%%%%%%%%%%%%%%%%%%%%%%%%%%%%%%%%%%%%%%%
%%%%%%%%%%%%%%%%%%%%%%%%%%%%%%%%%%%%%%%%

F.K. and B.V. gratefully acknowledge support through German Research Foundation (DFG) grant VO2413/2-1. 
E.M. and P.LF. were supported through NSF grant CBET-1638156. 
E.M. furthermore acknowledges support through U.S. Army ERDC grant W912HZ22C0037, U.S. ARO grant  W911NF-23-2-0046, and through NSF grant HS EAR 2100691.
The authors gratefully acknowledge the Gauss Centre for Supercomputing e.V. (www.gauss-centre.eu) for funding this project by providing computing time on the GCS Supercomputer SUPERMUC-NG at Leibniz Supercomputing Centre (www.lrz.de). 
The authors also gratefully acknowledge the computing time made available to them on the high-performance computer at the NHR Center of TU Dresden. This center is jointly supported by the Federal Ministry of Education and Research and the state governments participating in the NHR (www.nhr-verein.de/unsere-partner).

%%%%%%%%%%%%%%%%%%%%%%%%%%%%%%%%%%%%%%%%
%%%%%%%%%%%%%%%%%%%%%%%%%%%%%%%%%%%%%%%%
\section*{Author contributions}
%%%%%%%%%%%%%%%%%%%%%%%%%%%%%%%%%%%%%%%%
%%%%%%%%%%%%%%%%%%%%%%%%%%%%%%%%%%%%%%%%

P.LF. and B.V. designed and performed preliminary and main experiments.
P.LF. analyzed the experimental results.
F.K. performed and analyzed the numerical simulations.
F.K. and B.V. derived the empirical fit of the aggregate growth.
All authors discussed the results of the analyses.
F.K. generated the visualizations of the results.
F.K. prepared the manuscript.
All authors reviewed and revised the manuscript.

%%%%%%%%%%%%%%%%%%%%%%%%%%%%%%%%%%%%%%%%
%%%%%%%%%%%%%%%%%%%%%%%%%%%%%%%%%%%%%%%%
\section*{Competing interests}
%%%%%%%%%%%%%%%%%%%%%%%%%%%%%%%%%%%%%%%%
%%%%%%%%%%%%%%%%%%%%%%%%%%%%%%%%%%%%%%%%

The authors declare no competing interests.

%%%%%%%%%%%%%%%%%%%%%%%%%%%%%%%%%%%%%%%%
%%%%%%%%%%%%%%%%%%%%%%%%%%%%%%%%%%%%%%%%
\section*{Additional information}
%%%%%%%%%%%%%%%%%%%%%%%%%%%%%%%%%%%%%%%%
%%%%%%%%%%%%%%%%%%%%%%%%%%%%%%%%%%%%%%%%

\textbf{Supplementary information} The online version contains supplementary material.

\end{document}